\documentclass[12pt]{iopart}
\usepackage{iopams}
\usepackage[xdvi]{graphicx}

\newcommand{\bra}{\langle}
\newcommand{\ket}{\rangle}

\newcommand{\pder}[2]{\frac{\partial #1}{\partial  #2}}

\begin{document}


\title
[Thermodynamic transition associated with irregularly ordered 
ground states]
{Thermodynamic transition associated with irregularly ordered
ground states in a lattice gas model}

\author{Shin-ichi Sasa
\footnote[3]
{sasa@jiro.c.u-tokyo.ac.jp}}

\address{Department of Pure and Applied Sciences, University of Tokyo, 
Komaba, Tokyo 153-8902, Japan}

\begin{abstract}
A two-dimensional lattice gas model is proposed. 
The ground state of this model with a fixed density
is  neither periodic nor quasi-periodic. It also 
depends on system size in an irregular manner. 
On the other hand, it is ordered 
in the sense that the entropy density is zero in the 
thermodynamic limit. The existence of a thermodynamic 
transition associated with such irregularly ordered 
ground states is conjectured from a duality relation
for a thermodynamic function. This conjecture is 
supported by a phenomenological argument and  numerical 
experiments.
\end{abstract}





\section{Introduction}  %


The ground state of a crystal is described by the 
repetition of a unit cell. Recalling that the repetition 
is a simple deterministic rule generating an infinite size 
pattern from a finite size pattern, we become curious about 
equilibrium systems with ground states described by other 
deterministic rules than the repetition. Note that the 
entropy density of such deterministic ground states
is zero in the thermodynamic limit. In this sense,
the ground states are ordered as observed in crystals.
We call them {\it irregularly ordered ground states}. 
In this paper, we are concerned with thermodynamic 
transitions associated with irregularly ordered ground 
states.


We shall give  a precise definition of 
irregularly ordered ground states by considering
some examples.
First, quasi-crystals provide aperiodic ground 
states \cite{QC}.  However, since the Fourier 
transform of a quasi-crystal configuration consists 
of sharp delta peaks, we classify them as 
regular ground states, following the terminology in 
studies of dynamical systems \cite{DS,ER}. 
Another type is represented by ground states of 
the anti-ferromagnetic Ising model on a triangular 
lattice, which are often called disordered ones. 
The ground states are neither periodic nor quasi-periodic.
However, since their entropy density is not zero
as calculated explicitly \cite{Wannier},
we  do not regard such disordered ground states as 
irregularly ordered ground states. 
It is worthwhile noting that a string representation of
the ground states is  described by a stochastic rule 
corresponding to non-intersecting random walks 
\cite{Blote,Dhar}. 
Based on these considerations, we define 
irregularly ordered ground states as those
which have no delta peaks in their Fourier
transform and whose entropy density is zero. 

As an example with irregularly ordered 
ground states, one remembers the Frenkel-Kontorova 
model \cite{FK1,FK2}, because its ground state is
described by a chaotic map.
Since it does not exhibit a phase 
transition at finite temperature, it is natural 
to consider an extended Frenkel-Kontorova 
model in two-dimensions. Recently, for such an extended model 
together with other related models, a certain type of transition
has been found with the construction of ground states 
\cite{Yoshino}. An interesting study with experimental 
relevance is now going on. 
However, from a theoretical viewpoint, the model is still 
complicated if one attempts to elucidate the nature of 
thermodynamic transitions associated with irregularly 
ordered ground states. 
In order to develop 
a theoretical argument, it seems better to study a two-dimensional 
lattice model. We here remark that non-trivial ground states in 
lattice gas models, which include an infinitely series of periodic
ground states, were discussed in Refs. \cite{Kanamori,Dub}.
Furthermore, we should mention that Ref. \cite{Enter} reports
a three-dimensional short-range interaction model in which  
non-periodic long-range order sustains at positive temperature. 
This reference also provides a summary of mathematical results 
on non-periodic order of ground states and Gibbs states  
for lattice models. 


Based on these previous studies, in this paper, we first present 
a two-dimensional lattice gas  model with irregularly
ordered ground states. The ground states of our model are 
described by a one-dimensional cellular automaton whose 
deterministic rule yields a non-repetitive structure given 
by a superposition of Sierpinski gaskets. 
We next consider a possibility of thermodynamic transitions 
associated with the ground states. Since our model turns out 
to be self-dual for a duality transformation, we can present 
a simple argument for the thermodynamic transition.  

\section{Model} \label{model}  %



\begin{figure}[htbp]
\begin{center}
\begin{tabular}{cc}
\includegraphics[width=4cm]{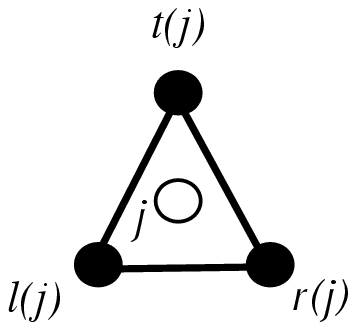}
\includegraphics[width=4cm]{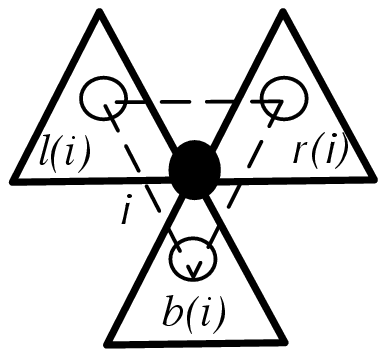}
\includegraphics[width=6cm]{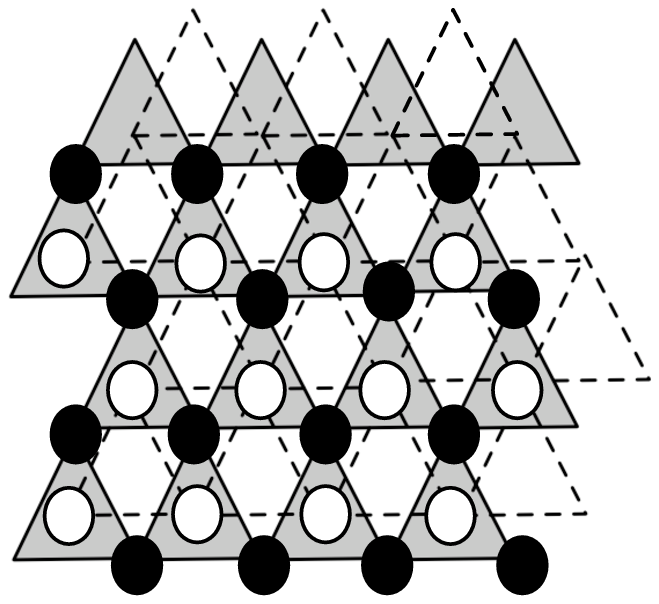}
\end{tabular}
\end{center}
\caption{ Filled circles represent sites in the original
triangular lattice $\Lambda$, and empty circles represent
sites in the dual lattice $\Lambda^*$.}
\label{fig0}
\end{figure}

We consider hard-core particles in a triangle lattice
consisting of $N$ sites. Let $\Lambda$  be the set of
all sites in the lattice and $\Lambda^*$ be the set 
of all upward triangles. 
We express a configuration of the particles by an occupation 
variable $\sigma_i$ for $i \in \Lambda$:  
$\sigma_i=1$ when site $i$ is occupied by a particle, 
while $\sigma_i=0$ 
when the site is empty. A particle configuration is collectively
denoted by  $\sigma=(\sigma_i)_{i \in \Lambda}$.  
For each upward triangle $j \in \Lambda^*$, 
we denote the sites at the top, left and right of 
the upward triangle by $t(j)$, $\ell(j)$, and $r(j)$, 
respectively. (See the left in figure \ref{fig0}.)
We  describe a soft interaction among particles 
in $j \in \Lambda^*$  by assuming a local energy $V_j$
as follows: $V_j(\sigma)=0$ when the particle number in $j$ is 
zero or two, otherwise  $V_j(\sigma)=J >0$. 
Explicitly, one may write
\begin{equation}
V_j(\sigma)=J[(\sigma_{t(j)}+\sigma_{\ell(j)}+\sigma_{r(j)})
\quad {\rm mod} \quad 2].
\label{V-1}
\end{equation}
Here, for a given site $i$, there are three upward triangles 
containing the site $i$. The upward triangles at the left, 
right and  bottom of the site $i$ are denoted by  $\ell(i)$, 
$r(i)$, and $b(i)$ $\in \Lambda^*$, respectively. (See the center in 
figure (\ref{fig0}).) We will see 
later that it is convenient to identify each upward triangle $j$ 
with a dual site taken at the center of the same triangle. Then, 
the set $\Lambda^*$ may be identified with the dual lattice, 
which also forms a triangular lattice.  See the right in 
figure~\ref{fig0}.

We study the equilibrium statistical mechanics with the Hamiltonian 
$H(\sigma)=\sum_{j \in \Lambda^*}V_j (\sigma)$. When studying statistical
mechanics, it is convenient to express 
\begin{equation}
H(\sigma)=\frac{J}{2}\sum_{j \in \Lambda^*}
[(2\sigma_{t(j)}-1) (2\sigma_{\ell(j)}-1)(2\sigma_{r(j)}-1)+1].
\label{Hamil}
\end{equation}
All the thermodynamic properties of the system with temperature
$T$ and density $\rho$ are described by the Helmholtz free 
energy density 
\begin{equation}
f(T,\rho)=- T \lim_{N \to \infty} \frac{1}{N}\log Z(T,\rho).
\end{equation}
Here, the partition function is given by
\begin{equation}
Z(T,\rho) =\sum_{\sigma} \e^{-\beta H(\sigma)}\delta(N\rho,\sum_{i}\sigma_i),
\label{Z:def}
\end{equation}
where $\delta(\ ,\ )$ is  Kronecker's delta function. 
Throughout the paper, $\beta$ is identical to $1/T$,
and the Boltzmann constant is set to unity. When we report
results of numerical experiments, we always assume $J=1$.
We also define the ground-state entropy explicitly
as follows. 
Let $\Omega(E,\rho;N)$ be the number of particle configurations 
for a given density $\rho$ and energy $E$. The entropy is defined 
by $S(E,\rho;N)=\log \Omega(E,\rho;N).$  By the ground-state entropy 
is meant $S_{\rm GS}(\rho;N)=\min_{E}S(E,\rho;N)$, 
and its density $s_{\rm GS}(\rho)$ is defined as 
$\lim_{N \to \infty} S_{\rm GS}(\rho;N)/N$. Obviously, 
$s_{\rm GS}(\rho) \ge 0$.
Note that the free energy density $f(T,\rho)$ is related to the entropy
as $f(T,\rho)= \lim_{N \to \infty} (E_*-TS(E_*,\rho;N))/N$,
where $E_*$ is determined by $ \partial S(E,\rho)/\partial E|_{E=E_*}
=\beta$.


The spin model equivalent to the Hamiltonian  (\ref{Hamil})
was introduced by Newman and Moore \cite{NM} and studied further 
\cite{GN, JG}. (The equivalence is explicitly
written in  (\ref{Y:spin}).) These previous studies 
focused on the spin system without  a magnetic field. 
As described later, 
this case corresponds to the lattice gas model with  zero 
chemical potential or with half-filled density. 
A thermodynamic transition, which is our main concern,
does not occur in this case.

\paragraph*{Ground states:}


The ground states of the model are defined as configurations 
in the limit $T \to 0$. Since $H \ge 0$, the configurations 
satisfying $H(\sigma)=0$ are ground-states. Obviously, $\sigma_i=0$ 
for any $i$ provides the ground state for the system 
with $\rho=0$. 
As the other limiting case, the maximally packed ground states 
are easily found. They are crystals with $\rho=2/3$. 
There are three configurations depending on its position. 


The other ground states  whose density is neither $0$ nor $2/3$
are given by non-trivial configurations. In order to see them,
let us consider the case that 
one particle is put on a site for the vacuum $\rho=0$. 
Then, the energy is 
non-zero in the three upward triangles containing this site.  
Thus, one particle must be put in each upward triangle so as 
to minimize the energy. This iterative process is not repetition, 
but it is given by a simple rule.


We describe this rule explicitly. From the expression ({\ref{V-1}),
one finds that the condition $V_j(\sigma)=0$ is equivalent 
to the addition rule $\sigma_{t(j)}=\sigma_{\ell(j)}+\sigma_{r(j)}$ (mod $2$). 
Suppose that
a particle configuration in the bottom line in the lattice is 
given. Then, when a particle configuration in the second bottom 
line is  determined by applying the addition rule, the energy
of upward triangles on the bottom line is zero. By repeating 
this process line by line, one obtains a particle configuration
in $\Lambda$. This deterministic rule is equivalent to an elementary 
cellular automaton (CA) on a one-dimensional lattice. 
According to Wolfram's classification, 
it corresponds to Rule 102 included in 
class III \cite{CA},  where space-time configurations 
generated by the rules in class III are known to be irregular 
for almost all initial conditions.


\begin{figure}[htbp]
\begin{center}
\begin{tabular}{cc}
\includegraphics[width=6cm]{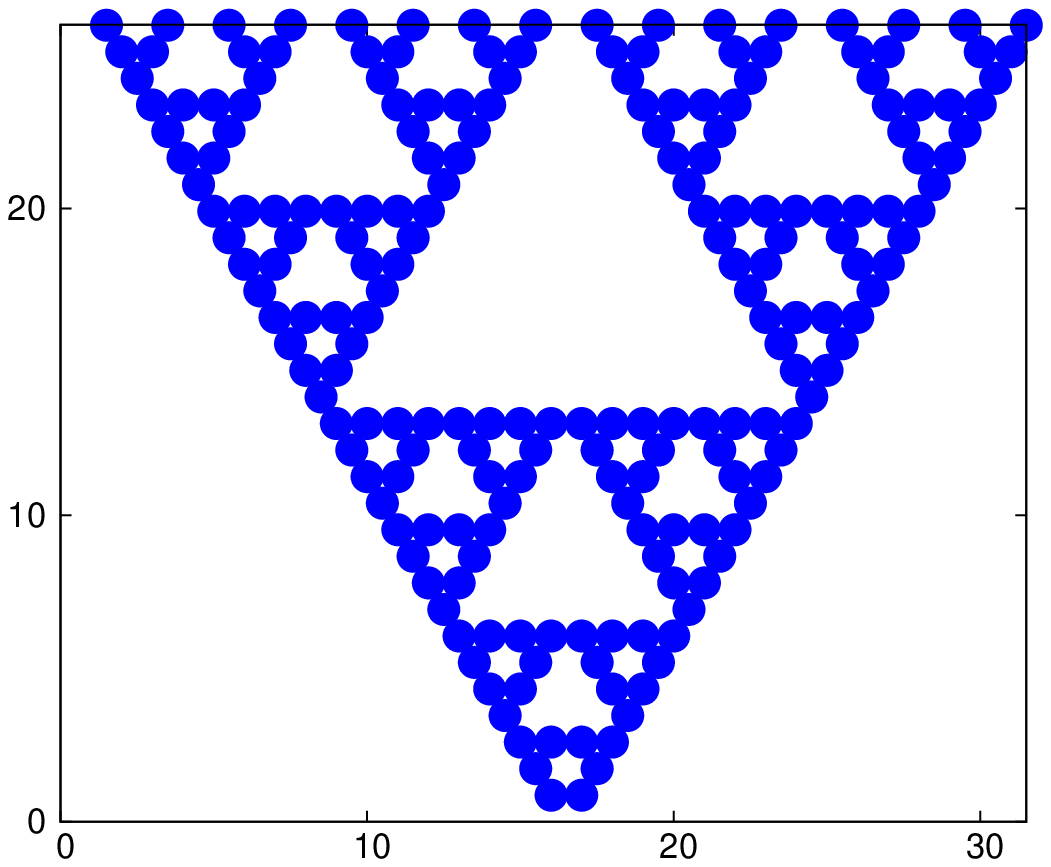}
\includegraphics[width=6cm]{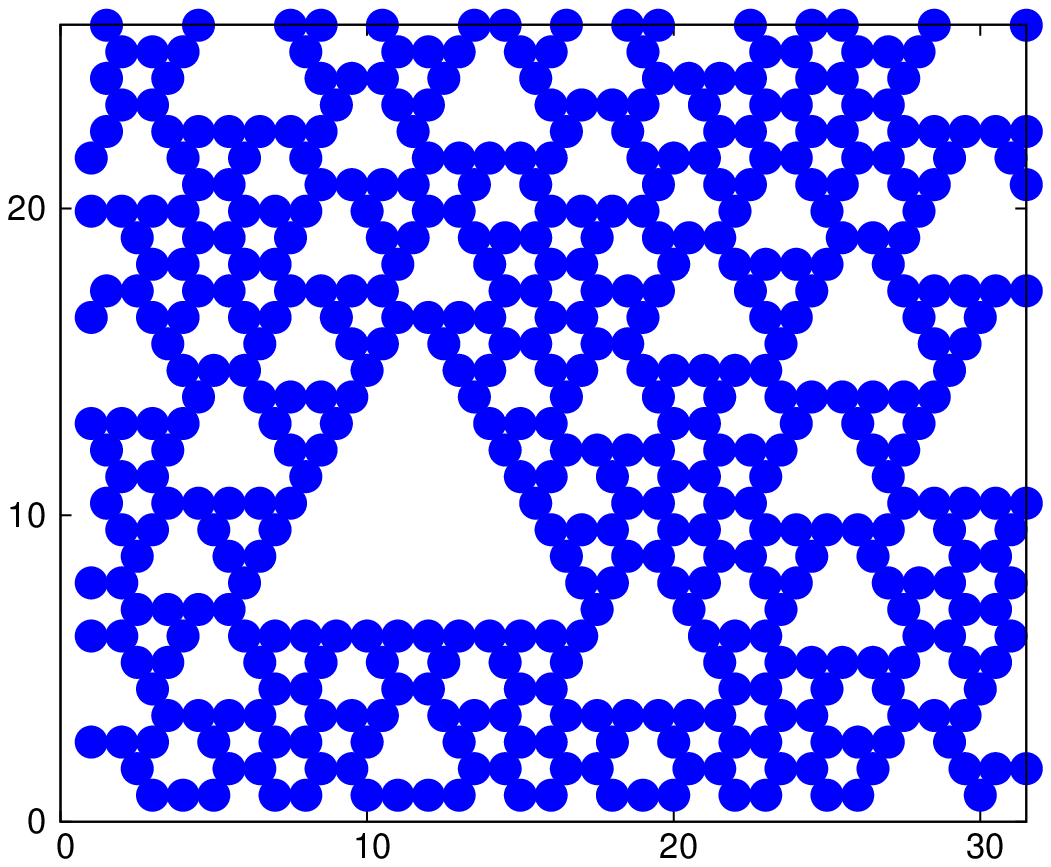}
\end{tabular}
\end{center}
\caption{(Color online) Space-time configuration of the cellular
 automaton. The blue circles represent sites at which 
$\sigma_i=1$. 
Periodic boundary conditions of the one-dimensional lattice
(horizontal axis) are assumed, and time is the vertical
axis. The initial conditions are
given by a single seed at the center of the bottom line
(left figure) and given randomly at the bottom line 
(right figure).}
\label{fig1}
\end{figure} 

As the simplest example, in the left of figure \ref{fig1},
we display a configuration starting from just one particle in 
the bottom line. It is called a Sierpinski gasket which is a 
famous  fractal pattern with the fractal dimension  $d_{\rm f}=
\log 3/\log 2$. 
Since the rule is linear, the superposition of two configurations 
(in the module two) also satisfies the CA rule. Thus, all the 
configurations generated by the CA rule are obtained by  superposition 
of Sierpinski gaskets.  
In the right of figure \ref{fig1},
we display a configuration starting from particles randomly
placed on the bottom line. 


Now, we consider the system with  linear size $L$. That is, $N=L^2$. 
For simplicity,
we first assume that the local energies for upward triangles
whose bottom edge is at the top line or whose top site
is at the bottom line are always zero.
We also assume periodic boundary conditions at the side lines. 
With this choice of boundary conditions, all the particle configurations 
satisfying $H(\sigma)=0$ are equivalent to  space time configurations 
generated by $L$ time steps evolution of the CA with the lattice 
size $L$. From the deterministic nature of the rule, the number of 
the configurations is $2^L$. Therefore, the density of 
ground-state entropy is estimated as $s_{\rm GS}(\rho) \le \log 2/L$
for any $\rho$, which leads to $s_{\rm GS}(\rho) =0$
in the thermodynamic limit.


For the other boundary conditions, to determine ground state 
configurations for a given $\rho$ is not easy. Concretely, 
let us consider the system with  periodic boundary conditions
in all directions. 
In principle, zero-energy configurations for a given system size 
$L$ can be determined by an algebraic method \cite{CA-algebra}. 
However, for example, it has been known that the number of the 
configurations depends on system size $L$ in a quite complicated 
manner which comes from the arithmetic nature of $L$. Furthermore, 
in some values of $L$, say $L=32$, there is no zero-energy configuration 
when $\rho \not = 0$. In this case, the ground state configuration 
involves defects, each of which is defined on an upward triangle
with the  energy $J$. 
The best positions of the defects 
for minimizing the energy would be determined in an irregular manner 
as a function of $\rho$ and $L$. 
Although the precise argument in this case is quite 
difficult, we provide a rough estimation of a typical number 
of defects $n_{\rm d}$ as follows. We start with a simple 
observation that an isolated Sierpinski gasket with any 
size $\ell$ is generated by assuming a defect at each vertex 
of the largest triangle. The number of particles involved 
in the gasket is $\ell^{d_{\rm f}}$. Let $n$ be the highest 
level of the gasket included in the system. Then, $2^n \simeq O(L)$.
When the particle density is finite, the number of gaskets is 
$O(L^{2-d_{\rm f}})$. Since there are three defects in
each gasket, $n_{\rm d} \simeq O(L^{2-d_{\rm f}})$. 
Therefore, the ground-state energy is $O(L^{2-d_{\rm f}})$
and the density of ground-state entropy is estimated as
$s_{\rm GS} \simeq O(\log 2/L)+O(n_{\rm d}/N) \simeq 
O(L^{-1})+O(L^{-d_{\rm f}})$, which leads to
$s_{\rm GS}(\rho)=0 $ in the thermodynamic limit.

\section{Statistical mechanics}\label{SM}   %

Following a standard method in statistical mechanics,
we consider the grand partition function 
\begin{equation}
\Xi(T,\mu) =\sum_{\sigma} \e^{-\beta [H(\sigma)-\mu \sum_{i}\sigma_i]}.
\label{Xi:def}
\end{equation}
The pressure $p(T,\mu)$ in the model is given by
\begin{equation}
p(T,\mu)= T \lim_{N \to \infty} \frac{1}{N}\log \Xi(T,\mu),
\end{equation}
and the free energy density $f(T,\rho)$ is obtained by 
the Legendre transformation
\begin{equation}
f(T,\rho)=\sup_{\mu}[\mu \rho-p(T,\mu)].
\label{p2f}
\end{equation}
Furthermore, by introducing a spin variable $s_i \equiv 2\sigma_i-1$ 
and parameters $K_1=\beta J/2 $ and $K_2= -\beta \mu/2$,  
we express  $\Xi(T,\mu)=\e^{-(K_1+K_2)N}Y(K_1,K_2)$ as 
\begin{equation}
Y(K_1,K_2) = 
\sum_{s} 
\e^{-K_1\sum_{j \in \Lambda^*}s_{t(j)} s_{\ell(j)}s_{r(j)}
-K_2\sum_{i\in \Lambda} s_i},
\label{Y:spin}
\end{equation}
where $s=(s_i)_{i \in \Lambda}$.
The partition function $Y$ is 
equivalent to that for the three-body Ising model on 
upward triangles under a magnetic field. Note that the 
case $K_2=0$ was studied in Refs. \cite{NM,GN,JG} and that
the three-body Ising model on all triangles without 
a magnetic field was solved exactly \cite{Baxter}.


We start our analysis from the easiest case 
$K_1 \gg 1 $ and $K_2 \simeq O(1)$, under which configurations 
minimizing 
$\sum_{j \in \Lambda^*}s_{t(j)} s_{\ell(j)}s_{r(j)}$ 
contribute to $Y$. 
We assume boundary conditions such that all the spins 
outside the system are down.  Then, the minimizer is uniquely
determined as $s_i=-1$ for all $i$. We thus obtain
\begin{equation}
Y(K_1,K_2)\simeq \exp( (K_1+K_2) N )
\label{low-1st}
\end{equation}
as the leading order contribution. 
This result leads to $p(T,\mu) \simeq 0$.
The first order correction to (\ref{low-1st})
arises from a single spin-flip with the probability 
$\exp(-3 \beta J)$. This means 
\begin{equation}
p(T,\mu)=O(\e^{-3\beta J}).
\label{p-ap}
\end{equation}

\paragraph*{Duality relation:}


Next we derive a duality relation for $p(T,\mu)$, which 
is presented in (\ref{dual-pressure}).
In the argument below, we assume $K_2 \ge 0$.
By making a change of variable $s_i \to -s_i$ in the sum of 
(\ref{Y:spin}), we have a relation
\begin{equation}
Y(K_1,K_2)=Y(-K_1,-K_2).
\label{sign-Y}
\end{equation}
With this relation, the expansion of the exponential function yields
\begin{equation}
\fl
Y(K_1,K_2) = C
\sum_{s}  
\left[\prod_{j \in \Lambda^*} 
(1+s_{t(j)} s_{\ell(j)}s_{r(j)} \tanh K_1)\right]
\left[\prod_{i \in \Lambda} (1+s_{i} \tanh K_2)\right],
\label{h-exp}
\end{equation}
where $C=(\cosh K_1)^N(\cosh K_2)^N$.
We then introduce a new variable 
$\tilde \sigma_j$ whose value is either 0 or 1
for $j \in \Lambda^*$.
Concretely, $\tilde \sigma_j=1$ 
when the term  $s_{t(j)} s_{\ell(j)}s_{r(j)} \tanh K_1$ is
picked up in the expansion of the first product. 
Otherwise, $\tilde \sigma_j=0$.
In terms of this new variable, the first product is rewritten
as 
\begin{eqnarray}
\fl
\prod_{j \in \Lambda^*} 
(1+s_{t(j)} s_{\ell(j)}s_{r(j)} \tanh K_1)
&=&
\sum_{\tilde \sigma} \prod_{j \in \Lambda^*} 
(s_{t(j)}s_{\ell(j)}s_{r(j)} \tanh K_1)^{\tilde \sigma_j} \nonumber \\
&=&
\sum_{\tilde \sigma} (\tanh K_1)^{\sum_{j \in \Lambda^*}\tilde \sigma_{j}}
\prod_{i \in \Lambda} 
s_{i}^{\tilde \sigma_{\ell(i)}+\tilde \sigma_{r(i)}+\tilde \sigma_{b(i)}},
\end{eqnarray}
where 
$\tilde \sigma=(\tilde \sigma_{j})_{j \in \Lambda^*}$
and  we have rearranged the product as that over $i\in \Lambda$.
Note that $\ell(i)$, $r(i)$, and $b(i)$ were defined previously.
See figure \ref{fig0}.
By substituting this expression into (\ref{h-exp}), 
we obtain
\begin{equation}
\fl
Y(K_1,K_2) =C
\sum_{s}  \sum_{\tilde \sigma} 
(\tanh K_1)^{\sum_{j \in \Lambda^*}\tilde \sigma_{j}}
\left[\prod_{i \in \Lambda} 
(s_{i}^{\tilde \sigma_{\ell(i)}+\tilde \sigma_{r(i)}+\tilde \sigma_{b(i)}}
+\tanh K_2 
s_{i}^{\tilde \sigma_{\ell(i)}+\tilde \sigma_{r(i)}+\tilde \sigma_{b(i)}+1})
\right].
\label{express}
\end{equation}


Here, we define $K_1^*$ by
\begin{equation}
\exp(-2 K_1^*)=\tanh K_2,
\label{dual-1}
\end{equation}
and introduce a variable $\tilde s_j=2\tilde \sigma_j-1$.
By  an explicit calculation, we  can confirm
\begin{equation}
\sum_{s_i}
(s_{i}^{\tilde \sigma_{\ell(i)}+\tilde \sigma_{r(i)}+\tilde \sigma_{b(i)}}
+\tanh K_2 
s_{i}^{\tilde \sigma_{\ell(i)}+\tilde \sigma_{r(i)}+\tilde \sigma_{b(i)}+1})
=
2 \e^{-K_1^* (\tilde s_{\ell(i)}\tilde s_{r(i)}\tilde s_{b(i)}+1)}.
\label{ident}
\end{equation}
Similarly, we write
\begin{equation}
(\tanh K_1)^{\sum_{j \in \Lambda^*}\tilde \sigma_{j}}
=\e^{-K_2^*\sum_{j \in \Lambda^*}(\tilde s_j+1)},
\label{ident-2}
\end{equation}
where $K_2^*$ has been defined by
\begin{equation}
\exp(-2 K_2^*)=\tanh K_1.
\label{dual-2}
\end{equation}
By substituting (\ref{ident}) and (\ref{ident-2}) into 
(\ref{express}), we obtain
\begin{equation}
Y(K_1,K_2) = 2^N C 
\e^{- K_2^*N^*}
\e^{- K_1^*N^*}
\sum_{\tilde s} 
\e^{- K_2^* \sum_{j \in \Lambda^*} \tilde s_{j}}
\e^{- K_1^* \sum_{i \in \Lambda }
\tilde s_{\ell(i)}\tilde s_{r(i)}\tilde s_{b(i)}},
\label{dual-3}
\end{equation}
where $\tilde s=(\tilde s_j)_{j \in \Lambda^*}$
and $N^*$ is the number of elements in $\Lambda^*$.
We have $N^*=N$ when we assume periodic boundary 
conditions. By recalling that $\Lambda^*$ may be 
identified with the dual lattice,
and by comparing (\ref{dual-3}) with (\ref{Y:spin}), 
one finds that the sum in (\ref{dual-3}) is nothing 
but the partition function of the same model with 
the new coupling constants $K_1^*$ and $K_2^*$ 
defined on the dual lattice $\Lambda^*$.
See figure \ref{fig0} again.
The only and irrelevant difference is that 
the particles now interact with each other 
on downward triangles.
We thus find the duality relation
\begin{equation}
Y(K_1,K_2) 
= 2^N (\cosh K_1)^N(\cosh K_2)^N 
\e^{-( K_1^*+K_2^*)N}Y(  K_1^*, K_2^*).
\label{dual-relation-1}
\end{equation}
We further simplify (\ref{dual-1}), (\ref{dual-2}),
and   (\ref{dual-relation-1}) as 
\begin{equation}
Y(K_1,K_2)= Y(K_1^*,K_2^*)(\sinh 2K_1 \sinh 2K_2)^{N/2},
\label{dual-relation-3}
\end{equation}
and   
\begin{equation}
(\sinh 2K_1^*)(\sinh 2K_2)=(\sinh 2K_2^*)(\sinh 2K_1)=1.
\label{dual-f}
\end{equation}
Finally, we define $T^*=1/\beta^*$ and $\mu^*$ by 
$K_1^*=\beta^* J/2 $ and $K_2^*= -\beta^* \mu^*/2$.
Then, (\ref{dual-relation-3}) yields
\begin{equation}
p(T,\mu)=p(T^*,\mu^*)-\frac{J}{2}
+\frac{\mu}{2}+T
\log \left(2\cosh \frac{J}{2T}\cosh \frac{\mu}{2T} \right).
\label{dual-pressure}
\end{equation}
Note that a duality relation connecting large $\beta$ to  
small $\beta\mu$ was discussed in Ref. \cite{Balian}.

\paragraph*{Exact result for $\rho=0.5$:}


As an application of the duality relation (\ref{dual-pressure}), 
we calculate the 
free energy density $f(T,\rho=0.5)$ exactly.  When $|\mu| \ll 1$, 
the dual point $T^* \ll 1$ satisfies $\exp(- J/T^*) \simeq -2 \mu/T$
for any  $T$. Since the pressure $p(T^*,\mu^*)$ in the regime 
$ T^* \ll 1 $ is known as (\ref{p-ap}), we substitute it 
into (\ref{dual-pressure}). 
Noting that $O(e^{-3\beta^* J})=O(|\mu|^3)$, we derive 
\begin{equation}
p(T,\mu)=-\frac{J}{2}
+\frac{\mu}{2}+T
\left[\log \left(2\cosh \frac{J}{2T} \right)
+O(|\mu|^2) \right]
\label{g}
\end{equation}
for $|\mu | \ll 1$. Since 
\begin{equation}
\rho=\left. \pder{p(T,\mu)}{\mu} \right\vert_{\mu=0}=\frac{1}{2},
\end{equation}
we obtain from (\ref{p2f})
\begin{equation}
f(T,\rho=0.5)=\frac{J}{2}
-T\left[\log \left( 2 \cosh \frac{J}{2T}\right)\right].
\end{equation}


This free energy density is equivalent to that of non-interacting 
Ising spins. Essentially the same result was obtained in
Ref. \cite{NM}.
Here, recall that the ground state of the system
with $\rho=0.5$ is given by superposition of the Sierpinski gaskets.
Therefore, in contrast to the simple form of the free energy density,
equilibrium configurations are complex as 
displayed in the left  of figure \ref{fig2}.
We then note that 
the equilibration time is extremely large, which might
originate from complex configurations in the system
with large $\beta$. 
In the right of figure \ref{fig2},
we plot the heat capacity obtained numerically with the aid of 
an exchange Monte Carlo method \cite{HukushimaNemoto}. 
The results for 
$\beta \ge 4$ still depend on the waiting time within our 
observations. 
This means that equilibrium states are not produced yet 
when $\beta \ge 4$. Indeed, the numerical data 
for $\beta \ge 4$ are different from the theoretical curve
derived exactly. One may interpret that the system 
is in a dynamical glass phase. See Refs. \cite{NM,GN,JG} 
for detailed studies of dynamical properties in this 
regime of the model. With regard to this problem, we 
also remark that the mean-field model of the ferro-magnetic 
three-body Ising model without a magnetic field exhibits 
a glass transition in a  meta-stable branch 
as well as a thermodynamic transition to 
a ferro-magnetic phase \cite{Franz}.

\begin{figure}[htbp]
\begin{center}
\begin{tabular}{cc}
\includegraphics[width=6cm]{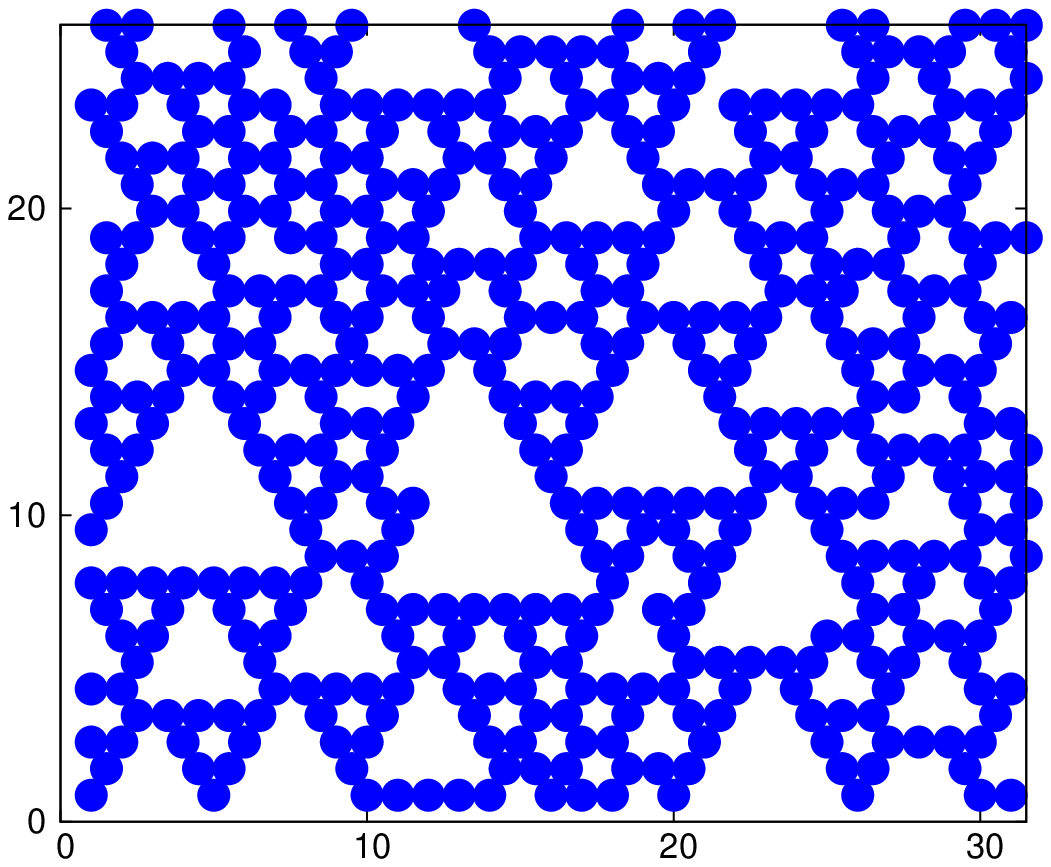}
\includegraphics[width=6cm]{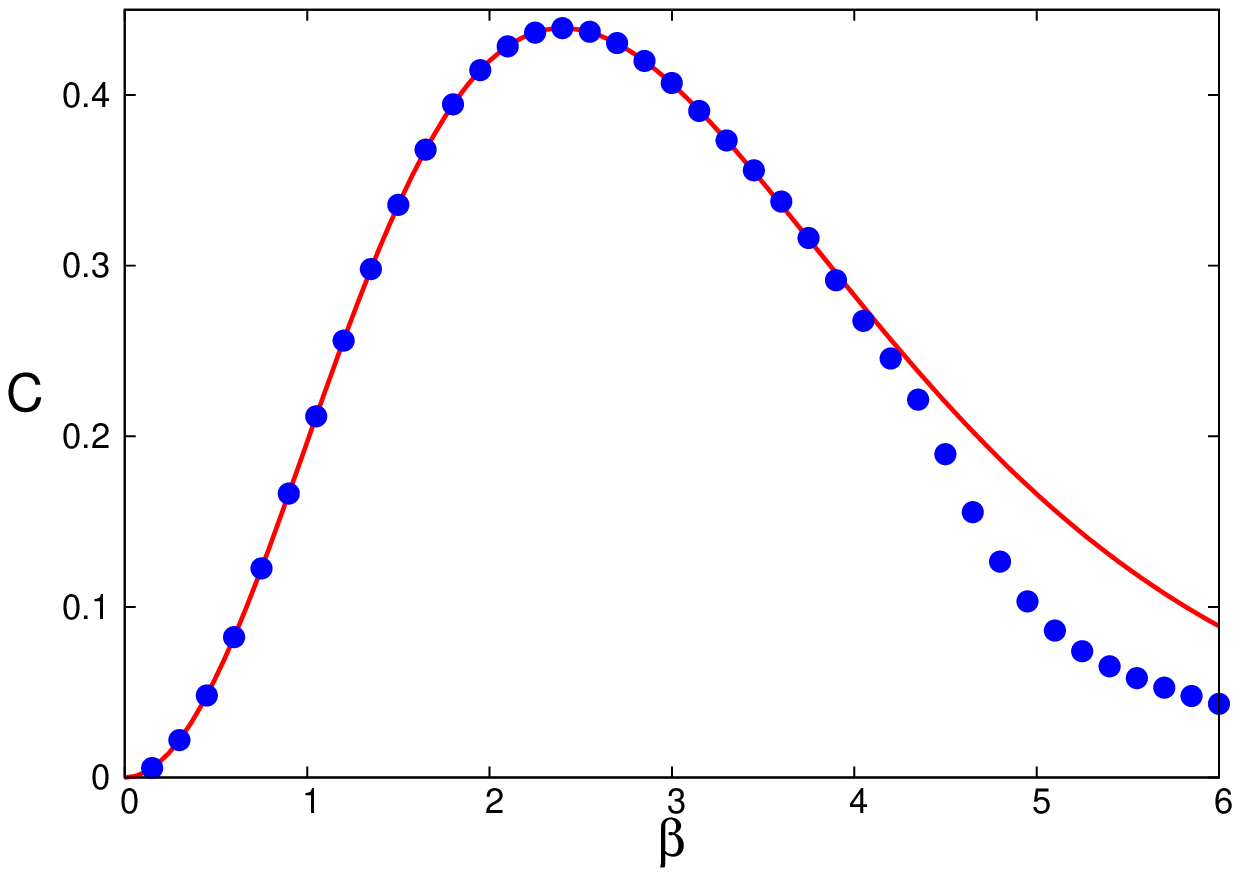}
\end{tabular}
\end{center}
\caption{
(Color online) 
 Numerical experiment of the model with $\mu=0$.
$N=1024$.
The left figure shows an equilibrium configuration 
for $\beta=3.75$. The blue circles represent 
particles. The right graph is the heat capacity 
$C$ as a function $\beta$. The circle symbols 
represent experimental results, where the data 
for $\beta \ge 4$ still depend on a waiting 
time within our observation. The solid line is 
the exact result derived theoretically.}
\label{fig2}
\end{figure} 

\paragraph*{Thermodynamic transition:}


The most standard application of duality relations is
to provide  candidates of thermodynamically singular
points. Indeed, if $p(T,\mu)$ is singular at some point
$(T_0,\mu_0)$, $p(T,\mu)$ is singular at $(T_0^*,\mu_0^*)$ too.
Thus, a thermodynamic transition occurs on the self-dual line
\begin{equation}
(\sinh J/T)(\sinh \mu/T)=-1
\label{self}
\end{equation}
if there is one transition point in $0 < T < \infty$ for fixed 
$\mu <0$. Note that, in general, there is  no reason to assume 
the latter condition, although there 
are many successful examples since the pioneering paper \cite{duality}.
As one example, let us consider the case where $T \gg 1$.
Then, the self-dual point is evaluated as $\mu \simeq -T\log T$.
It is hardly expected that a transition occurs in 
such a high-temperature and largely negative chemical
potential regime. Thus, without a hasty conclusion,
it is necessary to check whether or not transitions 
occur on the self-dual line.  


As another tractable case, we consider the system with 
$0 < -\mu/T \ll 1$ fixed. For sufficiently 
small $T$, we have $p(T,\mu)\simeq 0$ as derived  in 
(\ref{p-ap}). In this case, $\rho \simeq 0$. Now, we 
increase $T$ from 
the region $p(T,\mu)\simeq 0$ (with keeping $T \ll 1$)
and investigate the instability of the state $\rho\simeq 0$. 
Following a standard phenomenological argument \cite{Landau}, 
we replace a macroscopic spherical domain by a spherical domain 
whose pressure is given by (\ref{g}). The pressure difference 
from the bulk region is estimated by 
\begin{equation}
\Delta p \simeq \frac{\mu}{2}+T \exp(-J/T),
\end{equation}
where we have used (\ref{g}). If $\Delta p >0 $, the inserted 
domain grows if surface effects are ignored. 
That is, the first-order transition occurs at 
$\Delta p=0$, which leads to 
\begin{equation}
\frac{\mu}{2T}\exp( J/T) = -1.
\label{DeltaG-con}
\end{equation}
The equality coincides with the leading order equation  of 
(\ref{self}) under an asymptotic condition that $0 < -\mu/T \ll 1$  
and $T \ll 1$. Therefore, we expect
that the self-dual line under this condition provides 
actual transition points.  


\begin{figure}[htbp]
\begin{center}
\begin{tabular}{cc}
\includegraphics[width=6cm]{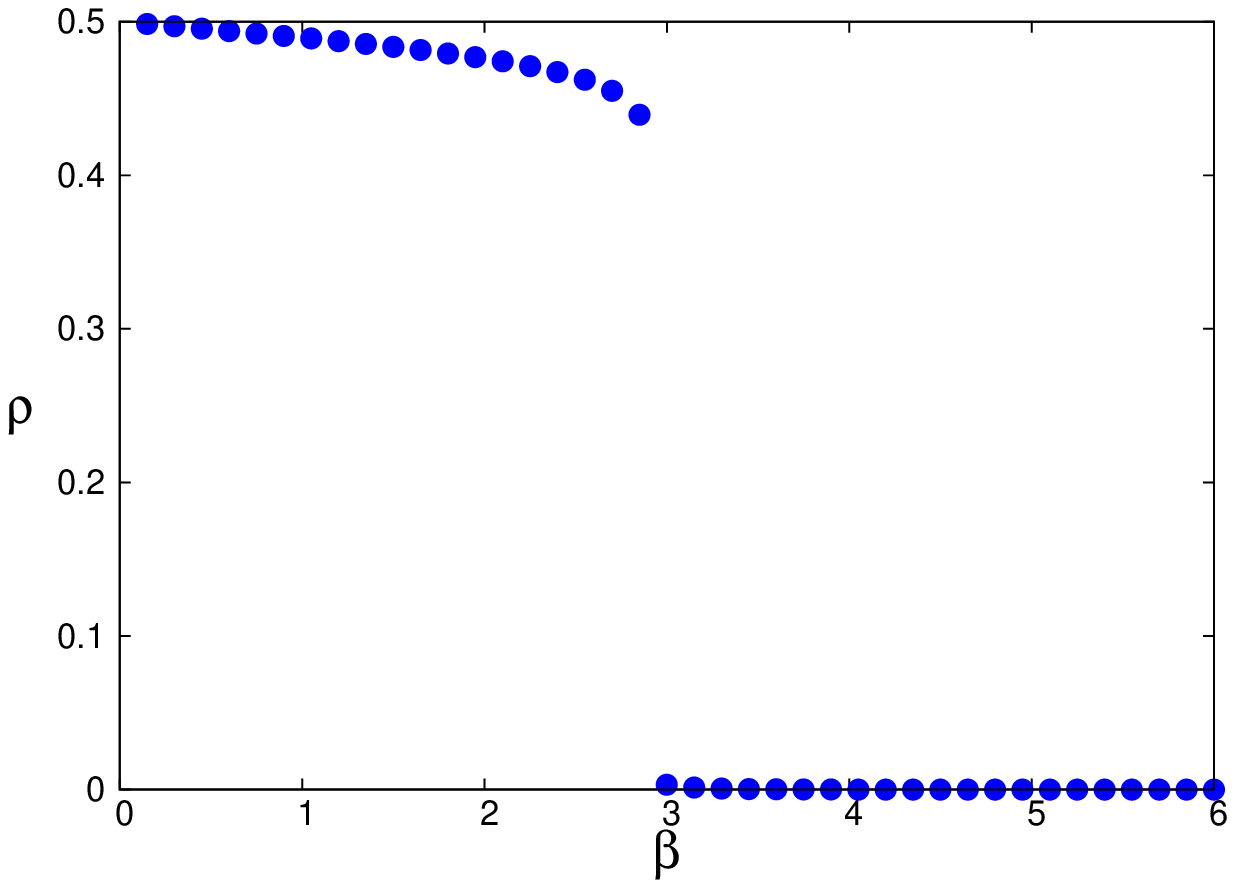}
\includegraphics[width=6cm]{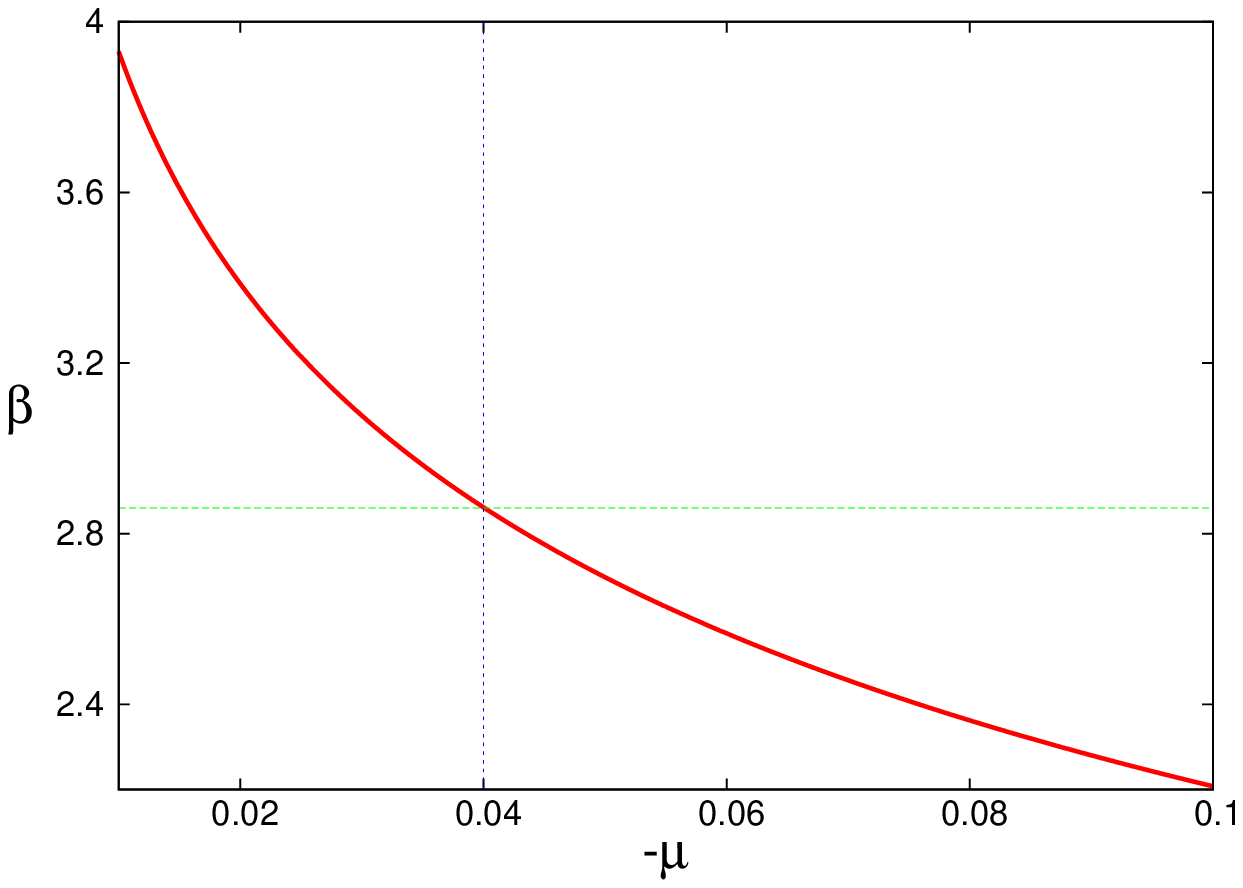}
\end{tabular}
\end{center}
\caption{
(Color online) 
 (Left): $\rho$ versus $\beta $ obtained 
by numerical experiments of the model with $\mu=-0.04$.
$N=1024$.
(Right): The self-dual relation 
given by (\ref{self}). Note that $\beta=2.86$ when
$\mu=-0.04$.
}
\label{fig3}
\end{figure} 

\begin{figure}[htbp]
\begin{center}
\begin{tabular}{cc}
\includegraphics[width=6cm]{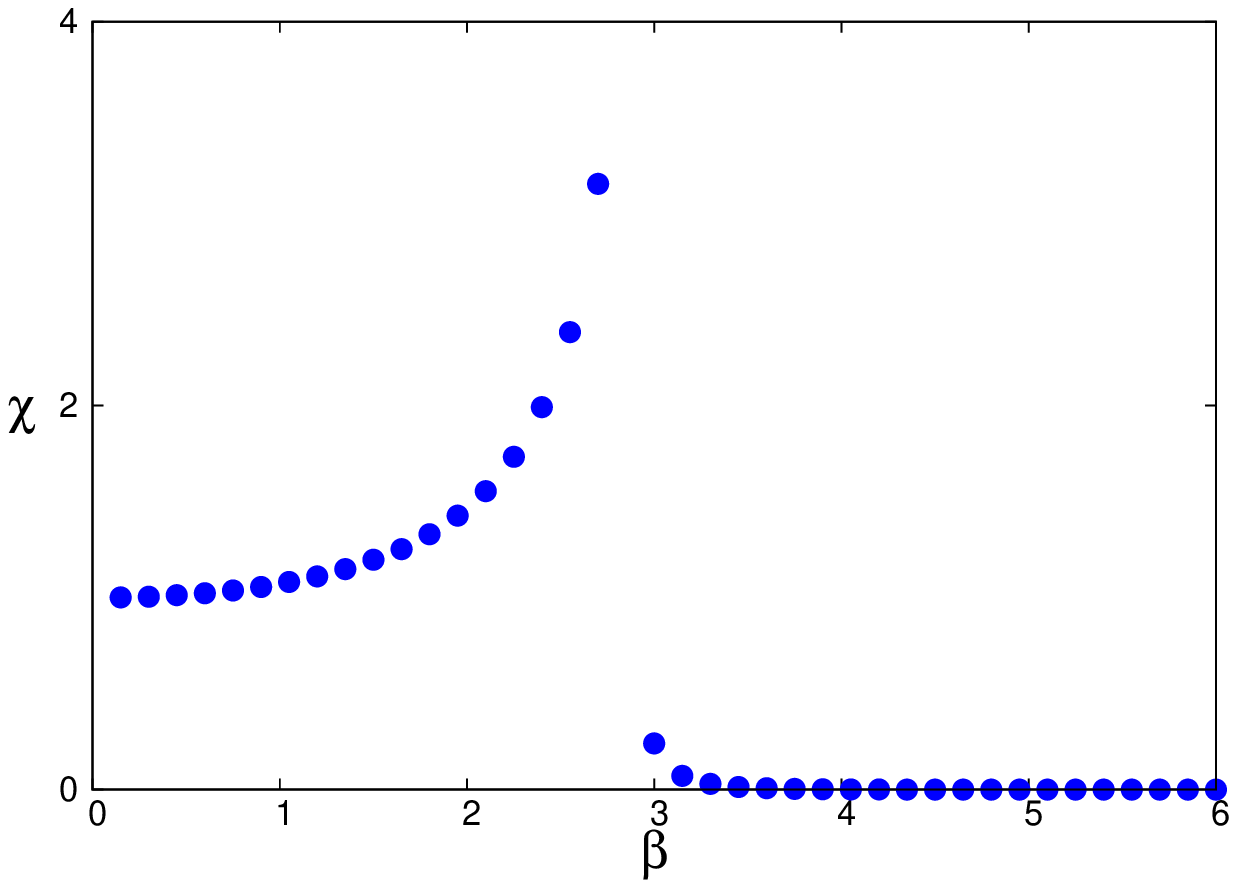}
\includegraphics[width=6cm]{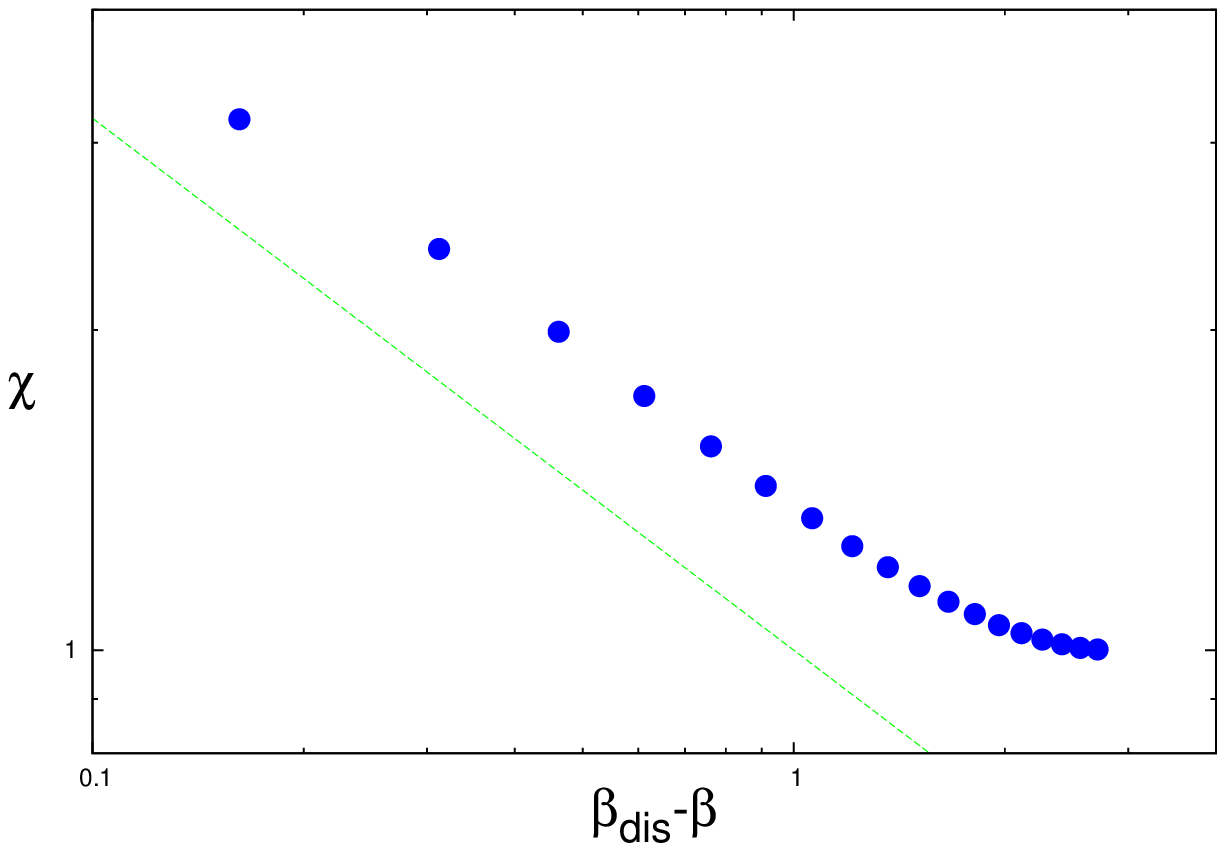}
\end{tabular}
\end{center}
\caption{
(Color online) 
 (Left): $\chi$ versus $\beta $ obtained 
by numerical experiments of the model with $\mu=-0.04$.
$N=1024$. (Right): Log-log plot of $\chi$ as a function
of $\beta_{\rm dis}-\beta$ for $\beta < \beta_{\rm dis}$.
The guide line represents  $\chi = 
(\beta_{\rm dis}-\beta)^{-1/2}$. 
}
\label{fig6}
\end{figure} 

In figure \ref{fig3},  we display a result of numerical 
experiments for the system with $\mu=-0.04$ fixed.
It is seen that the density exhibits the discontinuous
change from $\rho_1(\mu)$ to $\rho_2(\mu)$ 
at $\beta=\beta_{\rm dis}(\mu)$, where the left of 
figure \ref{fig3} shows that $2.75 < \beta_{\rm dis} <3.0$. 
Since the dual relation (\ref{self}) 
gives $\beta_{\rm dis} \simeq 2.86$ 
for $\mu=-0.04$, we judge that the theoretical consideration 
is consistent with the result of the numerical experiment. 
An  equilibrium 
configuration in the lower temperature phase 
$(T < T _{\rm dis}(\mu))$ is almost vacuum, while it is superposition 
of Sierpinski gaskets in the higher temperature phase 
$(T_{\rm dis}(\mu) < T \ll 1)$. 
Interestingly,
as shown in figure \ref{fig6}, the intensity of density 
fluctuations $\chi=N \bra (\rho-\bra \rho\ket)^2 \ket$ measured
with 
$(T,\mu)$ fixed exhibits a power-law behavior 
$\chi \simeq (\beta_{\rm dis}-\beta)^{-1/2}$ in the higher 
temperature side, while no divergent behavior in the 
lower temperature side. Similar behavior was observed
in several models \cite{Blote}.  
Furthermore, we note that the transition 
temperature $T_{\rm dis}(\mu)$ approaches  zero as $\mu \to 0 $. 
The equilibration becomes very hard as $|\mu| $ is decreased.
Indeed, we did not observe the first-order transition 
for the system with $\mu=-0.02$ for $N=1024$ within our
observation time.  In the left of figure \ref{fig5},
we present a schematic phase diagram in the $(\mu,T)$ 
space, where $\mu \le 0$. Since there is no phase 
transition for sufficiently large $T$, the first-order 
transition line in the state space possesses a terminal point
$(\mu_{\rm s},T_{\rm s})$.


\begin{figure}[htbp]
\begin{center}
\begin{tabular}{cc}
\includegraphics[width=6cm]{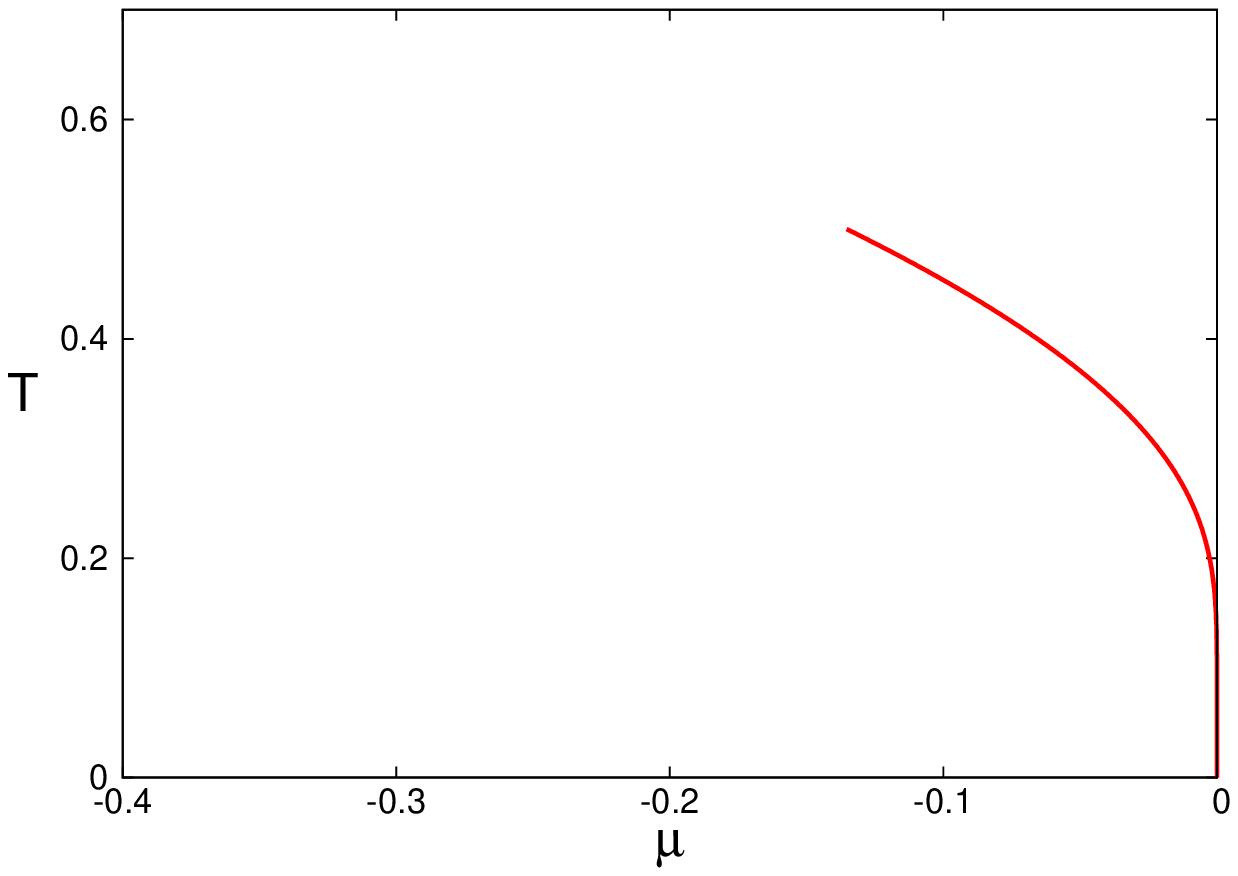}
\includegraphics[width=6cm]{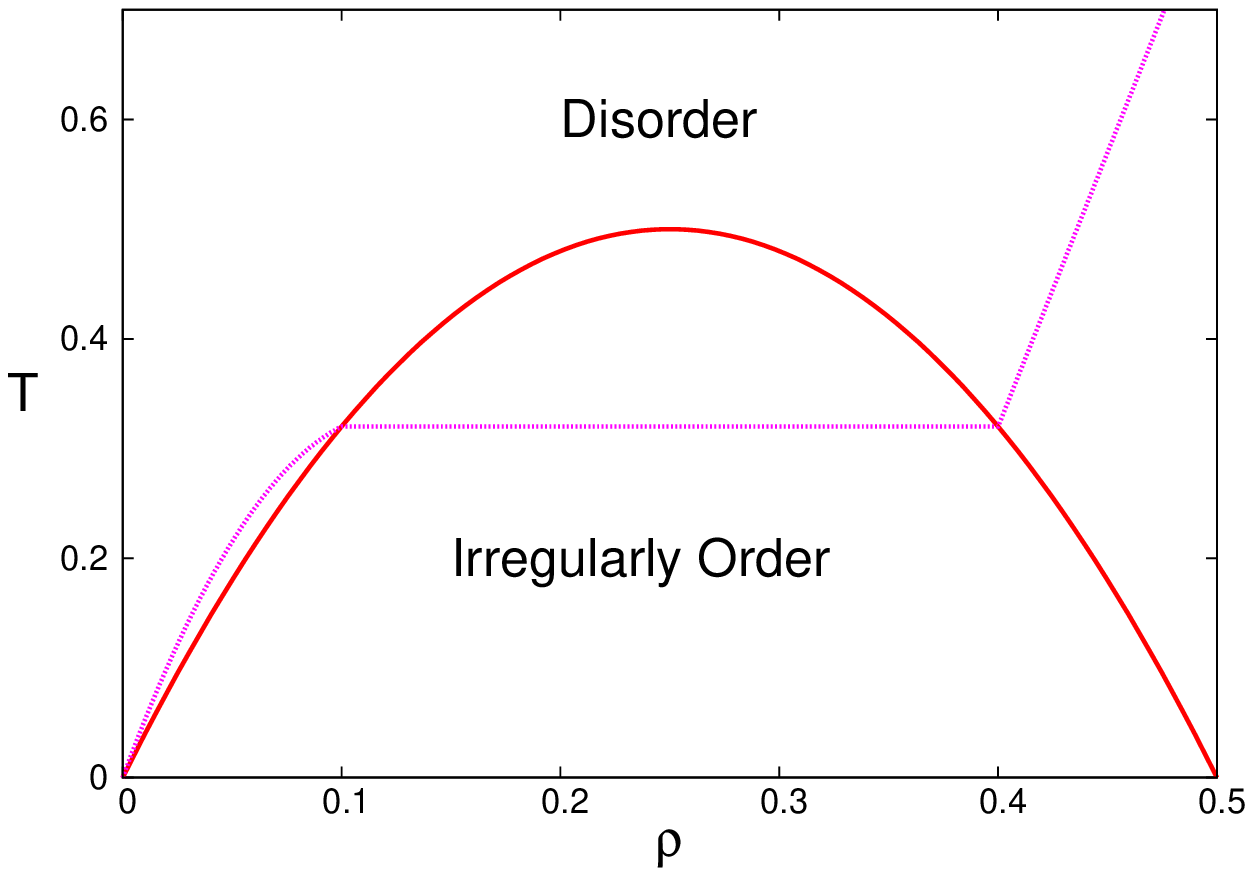}
\end{tabular}
\end{center}
\caption{
(Color online) 
Schematic phase diagrams in the state space $(\mu, T)$ (left)
and in the state space $(\rho, T)$ (right). In order to see
the qualitative behavior more easily, the graphs are not
accurately displayed. The chemical potential is equal to 
along the dotted line in the right figure. The coexistence
phase is labeled by ``Irregularly Order'',  but its 
non-trivial nature is not fully explored yet. 
See a discussion in the text. 
}
\label{fig5}
\end{figure} 

Let us consider a schematic phase diagram in the state space 
$(\rho,T)$. 
Since there exists a crystal configuration, a transition to 
a crystal state is expected to occur in the dense  regime with
$\rho >0.5$. Indeed, we observed it by a preliminary numerical 
experiment. However, in this paper, we restrict our investigation 
to the regime $\rho \le 0.5$ in order to present a clear argument.

We first recall that the density changes discontinuously from
$\rho_1(\mu)$ to $\rho_2(\mu)$ at $T_{\rm dis}(\mu)$ in decreasing
the temperature with $ \mu_{\rm s}< \mu <0$ fixed. 
See the dotted line in the right of figure \ref{fig5}. 
Then, the region
below the two curves $(\rho_1(\mu),T_{\rm dis}(\mu))$ and 
$(\rho_2(\mu), T_{\rm dis}(\mu))$, with $\mu_{\rm s} \le \mu \le 0$, 
is a coexistence phase that corresponds to the transition line
in the left of figure \ref{fig5}. 

Now, we fix $\rho$ satisfying $0 < \rho  < 1/2 $. There 
exists a special temperature $T_{\rm c}(\rho)$ below
which the system is in the coexistence phase. 
Note that $d{\mu}/d{\rho}|_{T}=0$ holds in this phase.
We further fix $T$ to be smaller than $T_{\rm c}(\rho) $. 
The chemical potential $\mu$ 
for a given $(T,\rho)$ is determined uniquely. 
Then, from the analogy with a liquid-gas transition, 
one may imagine  that 
two  phases, $(T, \rho_1(\mu))$ and $(T, \rho_2(\mu))$, 
are separated. Note that $\rho_1 \simeq 0$ and $\rho_2 \simeq 1/2$ 
when $T$ is sufficiently small　(that is, $\mu $ is 
sufficiently close to $0$).  Here, if this  picture is
correct, the interfacial energy proportional to $O(L)$
remains in the limit $T \to 0$.
However, as discussed in section \ref{model}, the 
ground states are  far from such simple configurations. 
At the end of section \ref{model}, we conjectured 
that point defects appear with the energy $O(L^{2-d_{\rm f}})$ 
in general ground states. Since this is less than the interfacial
energy $O(L)$, we expect that  phase separated 
configurations are not realized in the ground states.

Of course, even if phase separated configurations do not
appear in the ground states, this does not ensure the non-trivial 
nature of the system with finite temperature. We need to 
investigate the finite temperature system directly. 
More explicitly, we conjecture that 
the number of pure states in the coexistence phase 
is not two, but continuously distributed with a 
parameterization of density. That is, for example, 
when observing a density just on 
the first-order transition point in the $(T,\mu)$ ensemble, 
its distribution 
function may form a continuous function, not a 
two-peaks function, in the thermodynamic limit. 
To prove this conjecture mathematically is a challenging
problem, but careful numerical investigations will be done 
in future. With expecting the validity of our conjecture,
we call the coexistence phase  {\it irregularly ordered phase}.

\section{Future problems} \label{remark}   %


Before ending this paper, we list up  problems
to be studied in future. 
First of all, a precise phase diagram should be determined. 
In order to complete it, we will study the behavior near 
the critical point and the transition lines in detail.
After the determination of the phase boundaries, we will
characterize each phase. Among the phases, the most 
non-trivial is the irregularly ordered phase. 
As an example, in figure \ref{fig4}, 
we present configurations for $\rho=0.25$. The case 
$\beta=3.0$ (left) is in the disordered phase, 
while the case $\beta=4.5$ (right) is in the irregularly 
ordered phase. Both the configurations were obtained
by long-time calculation of an exchange Monte Carlo method.
The waiting time dependence is relatively small,
but more careful checks are necessary to judge whether 
the system reaches the equilibrium state in the 
irregularly ordered phase. 
Numerical studies on 
this phase are quite  challenging.

\begin{figure}[htbp]
\begin{center}
\begin{tabular}{cc}
\includegraphics[width=6cm]{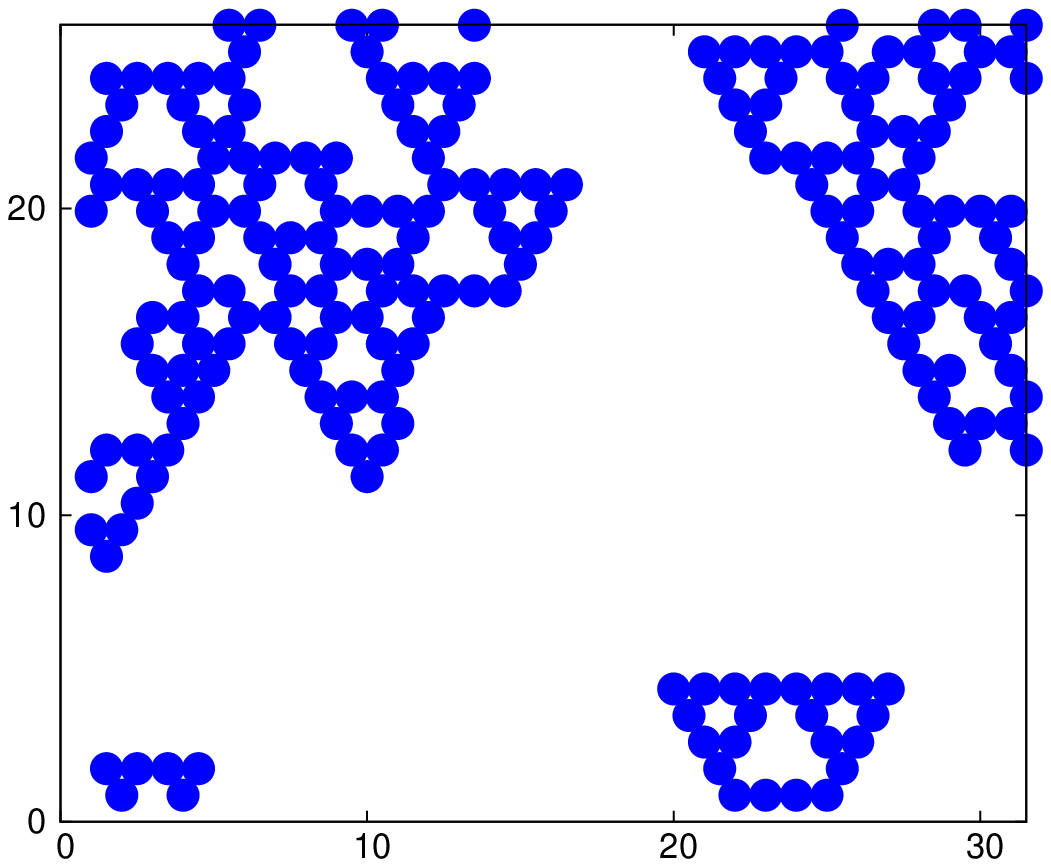}
\includegraphics[width=6cm]{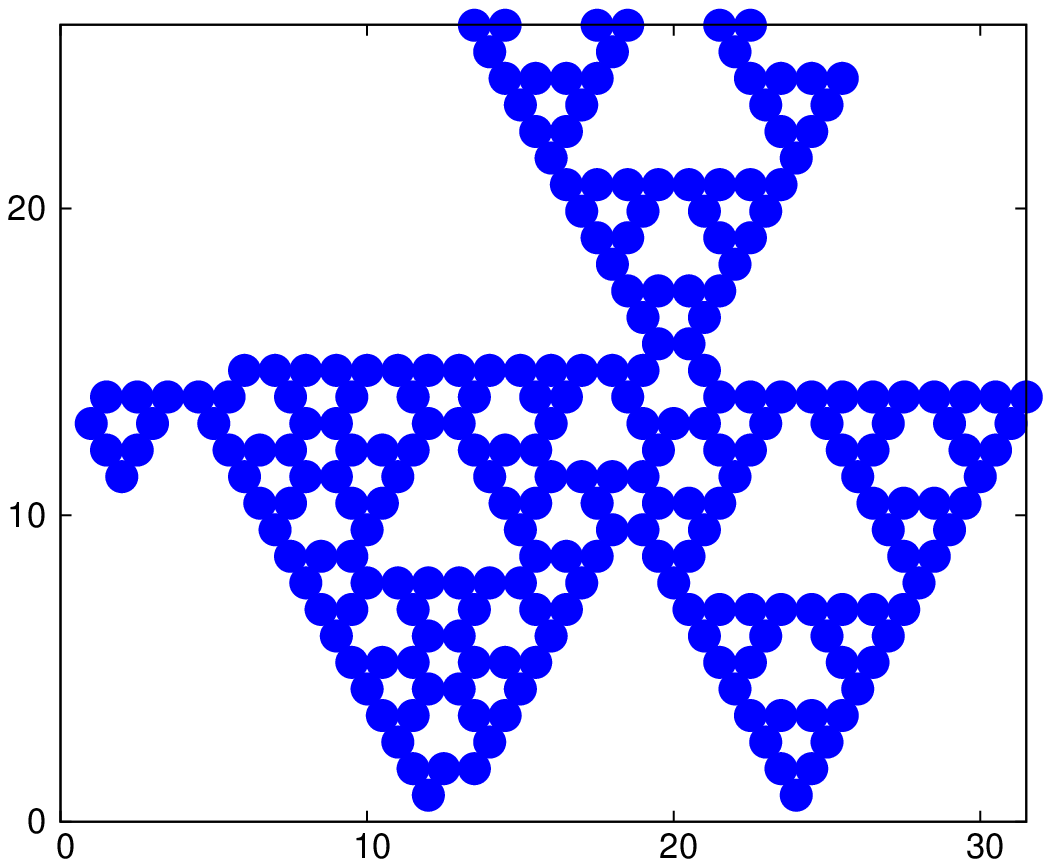}
\end{tabular}
\end{center}
\caption{
(Color online) 
Configurations for the system with $\rho=0.25$
fixed. The blue circles represent particles. 
$\beta=3.0$ (Left) and $\beta=4.5$ (Right)}
\label{fig4}
\end{figure} 


Theoretically, the next step study is 
to construct a perturbation theory for the system with 
small $T$ and small $|\mu|$ in order to understand
peculiar features observed numerically.  This
study will formulate calculation techniques for treating 
excitations of fractal patterns.  Here let 
us recall that the mapping to a free gas of defects
was useful to understand behaviors for the system
with $\mu=0$ \cite{GN,JG}. It is then a natural 
idea to describe the small $|\mu|$ system through 
a week interaction among defects. For example,
it might be interesting if one confirms that defects 
form a bounded state in the irregularly ordered state.
Furthermore, the study on algebraic properties 
of the cellular automaton rule may provide a hint for
characterizing the irregularly ordered phase.
As a different direction of study, one may investigate 
the irregularly ordered phase from a viewpoint of 
thermodynamic glass. At present, the existence of
thermodynamic glass in  finite dimensions is 
not established yet, 
mainly because  dynamical singularities cover the equilibrium
nature. Nevertheless, the understanding at the mean-field 
level has been greatly accumulated. In particular,  there
is consensus that thermodynamic glasses at the mean-field 
level are characterized by a one-step replica-symmetry 
breaking \cite{ParisiZamponi,MezardParisi}. 
Recently, a thermodynamic glass phase in a 
lattice model \cite{LatticeGlass}
has been investigated numerically 
along with this idea \cite{Parisi}.
As did in the study, one might argue
a replica-symmetry breaking in the 
irregularly ordered phase of our model.



Finally, we remark that  our model has no direct correspondence 
with experimental systems. However, needless to say, 
it is quite amazing to find  phase transitions associated with  
irregularly ordered ground states in natural phenomena. 
(See Ref. \cite{Barkema} for some connection of the model
to experimental systems.)
We expect that the phenomenon reported in this paper 
will be studied further theoretically and experimentally.


\ack

The author thanks A. C. D. van Enter, K. Hukushima, 
T. Nogawa, M. Ohzeki,  H. Tasaki and H. Yoshino 
for useful communications. This work was supported 
by grants from the Ministry of Education, 
Culture, Sports, Science, and Technology 
of Japan, Nos. 21015005 and 22340109.

\end{document}